\pdfoutput=1

\documentclass[11pt,twoside,a4paper,cmspaper,final,collab]{cms-tdr}
\def\svnVersion{475417P}\def\cmsCernNoTag{CERN-EP-2018-223}\def\cmsCernDate{\today}\def\cmsMessage{Published in Physical Review Letters as \href{http://dx.doi.org/10.1103/PhysRevLett.121.121801}{\doi{10.1103/PhysRevLett.121.121801.}}}

\begin{document}\cmsNoteHeader{HIG-18-016}

\hyphenation{had-ron-i-za-tion}
\hyphenation{cal-or-i-me-ter}
\hyphenation{de-vices}
\RCS$HeadURL: svn+ssh://svn.cern.ch/reps/tdr2/papers/HIG-18-016/trunk/HIG-18-016.tex $
\RCS$Id: HIG-18-016.tex 474778 2018-09-12 14:54:34Z jolsen $

\newlength\cmsFigWidth
\ifthenelse{\boolean{cms@external}}{\setlength\cmsFigWidth{0.49\textwidth}}{\setlength\cmsFigWidth{0.6\textwidth}}
\newlength{\cmsTabSkip}\setlength\cmsTabSkip{1ex}

\newcommand{\Vvar}{\ensuremath{\cmsSymbolFace{V}}\xspace}
\newcommand{\HBB}{\ensuremath{\PH\to\bbbar}\xspace}
\newcommand{\mH}{\ensuremath{m_\PH}\xspace}
\newcommand{\mjj}{\ensuremath{m\mathrm{(jj)}}\xspace}
\newcommand{\ptjj}{\ensuremath{{\pt\mathrm{(jj)}}}\xspace}
\newcommand{\ptll}{\ensuremath{{\pt}(\ell\ell)}\xspace}
\newcommand{\ptV}{\ensuremath{\pt(\Vvar)}\xspace}
\newcommand{\twol}{2-lepton\xspace}
\newcommand{\onel}{1-lepton\xspace}
\newcommand{\VH}{\ensuremath{{\Vvar}\PH}\xspace}
\newcommand{\VZ}{\ensuremath{{\Vvar}\cPZ}\xspace}
\newcommand{\WH}{\ensuremath{\PW\PH}\xspace}
\newcommand{\WZ}{\ensuremath{\PW\cPZ}\xspace}
\newcommand{\ZBB}{\ensuremath{\cPZ\to\bbbar}\xspace}
\newcommand{\zerol}{0-lepton\xspace}
\newcommand{\ZH}{\ensuremath{\cPZ\PH}\xspace}
\newcommand{\Vjets}{\ensuremath{\Vvar}+jets\xspace}

\cmsNoteHeader{HIG-18-016}
\title{Observation of Higgs boson decay to bottom quarks}

\date{\today}

\abstract{
The observation of the standard model (SM) Higgs boson decay to a pair of bottom quarks is presented. The main contribution to this result is from processes in which Higgs bosons are produced in association with a \PW\ or \cPZ\ boson (\VH), and are searched for in final states including 0, 1, or 2 charged leptons and two identified bottom quark jets. The results from the measurement of these processes in a data sample recorded by the CMS experiment in 2017, comprising 41.3\fbinv of proton-proton collisions at $\sqrt{s} = 13\TeV$, are described. When combined with previous \VH measurements using data collected at $\sqrt{s}=7$, 8, and 13\TeV, an excess of events is observed at $\mH = 125\GeV$ with a significance of 4.8 standard deviations, where the expectation for the SM Higgs boson is 4.9.  The corresponding measured signal strength is $1.01\pm 0.22$. The combination of this result with searches by the CMS experiment for \HBB in other production processes yields an observed (expected) significance of 5.6 (5.5) standard deviations and a signal strength of $1.04\pm 0.20$.
}

\hypersetup{%
pdfauthor={CMS Collaboration},%
pdftitle={Observation of Higgs boson decay to bottom quarks},%
pdfsubject={CMS},%
pdfkeywords={CMS, physics, Higgs}}

\maketitle

Since the discovery of a new boson with a mass near 125\GeV by the ATLAS~\cite{Aad:2012tfa} and CMS~\cite{Chatrchyan:2012ufa,Chatrchyan:2013lba} Collaborations, rapid progress in the understanding of its properties and couplings has revealed that the new particle is compatible with the standard model (SM) Higgs boson \PH~\cite{Englert:1964et,Higgs:1964ia,Higgs:1964pj,Guralnik:1964eu,Higgs:1966ev,Kibble:1967sv}. Observation of Higgs boson decays in the $\gamma \gamma$, $\cPZ\cPZ$, $\PW\PW$, and $\Pgt\Pgt$ modes have been reported~\cite{Aad:2014eha,Khachatryan:1728107,Aad:2014eva,Chatrchyan:2013mxa,ATLAS:2014aga,Aad:2015ona,Chatrchyan:1633401,Aad:2015vsa,Chatrchyan:1643937,Sirunyan:2273798,Sirunyan:2018egh}, and all measured properties~\cite{Aad:2015gba,Khachatryan:2014jba,Chatrchyan:2012jja,Aad:2013xqa,Khachatryan:2016vau,Aad:2015zhl,Sirunyan:2017exp,Aaboud:2018ezd,Aaboud:2018wps} support this hypothesis. Recently, the CMS and ATLAS Collaborations reported observations of the Higgs boson produced in association with a top quark pair that are compatible with the SM prediction, representing the first direct measurements of the Higgs boson coupling to quarks~\cite{Sirunyan:2018hoz,Aaboud:2018urx}.

The decay $\HBB$, with a predicted branching fraction of about 58\%~\cite{deFlorian:2016spz} for a Higgs boson mass of $\mH=125\GeV$, is the most prevalent decay mode but has not yet been established unequivocally. A precise measurement of the rate for this process directly probes the Yukawa coupling of the Higgs boson to a down-type quark, and provides a necessary test of the hypothesis that the Higgs field is the source of mass generation in the charged fermion sector of the SM~\cite{PhysRevLett.19.1264,Nambu:1961fr}. At both the LHC and the Tevatron the most sensitive production process in the search for $\HBB$ decays is when the Higgs boson is produced in association with a vector boson (\VH). The CDF and D0 Collaborations at the Tevatron reported an excess of events in this channel with a significance of 2.8 standard deviations ($\sigma$) at a mass of $\mH = 125\GeV$~\cite{PhysRevLett.109.071804}. Last year, the ATLAS and CMS Collaborations reported evidence for the \VH, \HBB\ process at a mass of $\mH = 125\GeV$ corresponding to observed (expected) significances of 3.6 (4.0)$\sigma$ and 3.8 (3.8)$\sigma$, respectively, combining data collected during Run 1 at $\sqrt{s}=7$ and 8\TeV, and Run 2 at 13\TeV~\cite{Aaboud:2017xsd,Sirunyan:2017elk}. Searches for the $\HBB$ decay in other production processes, with less sensitivity than \VH, have also been reported at the LHC~\cite{Sirunyan:2017dgc,Aad:2015gra,Aad:2016zqi,Chatrchyan:2013yea,Khachatryan:2014qaa,Aaboud:2016cns,Khachatryan:2015bnx}.

In this Letter we present an observation of the Higgs boson decay to bottom quarks. The measurement described here examines the $\VH$ production process, where the Higgs boson is produced in association with a $\PW$ or $\cPZ$ boson and decays into $\cPqb\cPaqb$. The data comprise proton-proton ($\Pp\Pp$) collisions recorded at $\sqrt{s}=13\TeV$ by the CMS detector at the LHC in 2017, corresponding to a total integrated luminosity of $41.3\fbinv$~\cite{CMS-PAS-LUM-17-004}. Five distinct final states are considered: $\cPZ(\nu\nu)\PH$, $\PW(\mu\nu)\PH$, $\PW(\Pe\nu)\PH$, $\cPZ(\mu\mu)\PH$, and $\cPZ(\Pe\Pe)\PH$, corresponding to three {\it channels} with $0$, $1$, or $2$ charged leptons from the vector boson decay. In addition, two identified jets due to hadronization of bottom quarks (\cPqb\ jets) from the Higgs boson decay are required. Important background processes include the production of \PW\ and \cPZ\  bosons in association with jets (\Vjets), production of top quark pairs (\cPqt\cPaqt) and single top quarks (\cPqt), diboson (\PW\PW, \PW\cPZ, \cPZ\cPZ), and multijet (QCD) events.

The analysis presented here closely follows the methods previously used to search for the \VH, \HBB process~\cite{Sirunyan:2017elk} and incorporates several improvements, including more efficient identification of \cPqb\ jets, better dijet mass resolution, and the use of different multivariate discriminant techniques that better separate signal from background.  For each channel, a signal region enriched in \VH events is selected together with several control regions, each enriched in events from individual background processes. A simultaneous binned-likelihood fit to the shape and yield (normalization) of specific distributions for the signal and control regions for all channels combined is used to extract a possible Higgs boson signal. To validate the procedure, the same methodology is used to extract a signal for the associated production process \VZ, with \ZBB, which has an identical final state to \VH with \HBB, but with a production cross section 5 to 15 times larger, depending on the kinematic regime considered. The result from this measurement is combined with those of similar measurements performed by the CMS Collaboration using 2016 data and Run 1 data, in the \VH and other production processes. This combination leads to an observation of the Higgs boson decay to bottom quarks. An observation of this decay mode by the ATLAS Collaboration~\cite{Aaboud:2018zhk} was submitted for publication at the same time as this Letter.

The central feature of the CMS apparatus is a superconducting solenoid of 6\unit{m} internal diameter, providing a magnetic field of $3.8\unit{T}$. A silicon pixel and strip tracker, a lead tungstate crystal electromagnetic calorimeter, and a brass and scintillator hadron calorimeter, each composed of a barrel and two endcap sections, reside within the solenoid. Forward calorimeters extend the pseudorapidity coverage provided by the barrel and endcap detectors. Muons are detected in gas-ionization chambers embedded in the steel flux-return yoke outside the solenoid. A more detailed description of the CMS detector, together with a definition of the coordinate system and the relevant kinematic variables, can be found in~\cite{Chatrchyan:2008aa}.

Signal and background processes are simulated with several Monte Carlo (MC) event generators, while the CMS detector response is modeled with \GEANTfour~\cite{GEANT4}. The quark-induced \ZH and \WH signal processes are generated at next-to-leading order (NLO) QCD accuracy using the {\POWHEG} v2~\cite{Nason:2004rx,POWHEG,Alioli:2010xd} event generator extended with the MiNLO procedure~\cite{Hamilton2012,Luisoni:2013kna}, while the gluon-induced \ZH process is generated at leading order (LO) accuracy with {\POWHEG} v2. The Higgs boson mass is set to $125\GeV$ for all signal samples. Diboson background events are generated with {\MGvATNLO} v2.4.2~\cite{Alwall:2014hca} at NLO with the FxFx merging scheme~\cite{Frederix:2012ps} and up to two additional partons. The same generator is used at LO accuracy with the MLM matching scheme~\cite{Alwall:2007fs} to generate \Vjets events in inclusive and b-quark enriched configurations with up to four additional partons, and to generate a sample of QCD events. The \ttbar~\cite{Frixione:2007nw} and single $\cPqt$ production processes in the \cPqt\PW~\cite{Re:2010bp} and t~\cite{Frederix:2012dh} channels are generated to NLO accuracy with {\POWHEG} v2, while the s channel~\cite{Alioli:2009je} single \cPqt\ process is generated with {\MGvATNLO} v2.4.2. The parton distribution functions used to produce all samples are the next-to-next-to-leading order (NNLO) {NNPDF3.1} set~\cite{Ball:2017nwa}. For parton showering and hadronization, the matrix element generators are interfaced with {\PYTHIA} v8.230~\cite{Sjostrand:2014zea}. For all samples, simulated additional $\Pp\Pp$ interactions (pileup) are added to the hard-scattering process with the multiplicity distribution matched to the 2017 data.

The production cross sections for the signal samples are rescaled as a function of the vector boson transverse momentum, \ptV, to NNLO QCD + NLO electroweak accuracy combining the \textsc{vhnnlo}~\cite{Ferrera:2013yga,Ferrera:2014lca,Ferrera:2011bk,Ferrera:2017zex}, \textsc{vh@nnlo}~\cite{Brein:2012ne,Harlander:2013mla}, and \textsc{hawk} v2.0~\cite{Denner:2014cla} generators as described in~\cite{deFlorian:2016spz}. The production cross sections for the \ttbar samples are rescaled to the NNLO prediction with the next-to-next-to-leading-log result obtained from {{\textsc{Top++} v2.0}}~\cite{Czakon:2011xx}, while the \Vjets samples are rescaled to the NNLO cross sections using \FEWZ 3.1~\cite{Li:2012wna}. In the V+jets samples used in this analysis, the \ptV spectrum in data is observed to be softer than in simulated samples, as expected from higher-order electroweak corrections to the production processes~\cite{NLO_plus_QCD_weight}. Events in each channel are reweighted to account for electroweak corrections to \ptV, which reach up to 10\% for \ptV near 400\GeV. In addition, a differential LO-to-NLO correction is applied as a function of the separation in $\eta$ between the two jets from the Higgs boson decay~\cite{Sirunyan:2017elk}. The \ttbar samples are reweighted as a function of top quark \pt to account for the known difference between data and simulation~\cite{PhysRevD.95.092001}.

The CMS particle-flow (PF) event algorithm~\cite{CMS-PRF-14-001} is used to reconstruct and identify individual particles (PF objects) with an optimized combination of information from the various elements of the CMS detector. This algorithm is employed at the trigger level, and in the more detailed reconstruction of data that occurs offline. The objects identified by the algorithm comprise candidate electrons, muons, photons, and charged as well as neutral hadrons. Jets are reconstructed from PF objects using the anti-\kt clustering algorithm~\cite{Cacciari:2008gp} implemented in the FASTJET package~\cite{Cacciari:fastjet1,Cacciari:fastjet2}, with a distance parameter of 0.4. The missing transverse momentum vector $\ptvecmiss$ is defined as the negative vectorial \pt sum of all the PF objects identified in the event, and its magnitude is referred to as $\ptmiss$. 

The reconstructed vertex with the largest value of summed physics-object $\pt^2$ is considered to be the primary $\Pp\Pp$ interaction vertex. The physics objects used in this calculation are jets clustered using the jet finding algorithm with the tracks assigned to the vertex as inputs, and the associated missing transverse momentum computed as the negative vectorial \pt sum of those jets. All charged hadrons that originate from pileup are removed from consideration in the event. In addition, the average neutral energy density from pileup is evaluated from PF objects and subtracted from the reconstructed jets as well as from the summed energy in the vicinity of leptons (isolation), as described in~\cite{CMS-PRF-14-001}.

Events of interest are selected online using a two-tiered trigger system~\cite{Khachatryan:2016bia} based on custom hardware processors and a farm of commercial processors running a version of the full offline reconstruction software optimized for speed. Events in the \zerol channel are selected primarily by a trigger requiring both \ptmiss and \mht to be larger than 120\GeV, where \mht is defined as the magnitude of the negative vectorial \pt sum of all jets. Single-lepton triggers are used to select events in the \onel channel, where the leptons are required to be isolated from other PF objects. The single-muon trigger requires $\pt > 27\GeV$, while the single-electron trigger requires $\pt > 32\GeV$. For the \twol channel, dilepton triggers are employed with minimum \pt requirements of $(17,8)\GeV$ for the two muons and $(23,12)\GeV$ for the two electrons. After offline requirements, the \zerol trigger efficiency reaches 100\% above $\ptmiss\sim 220\GeV$, while the \onel triggers have efficiencies of approximately 95\% for muons and 90\% for electrons, and the dilepton trigger efficiency is approximately 91\% for muons and 96\% for electrons.

Events are selected offline based on the presence of 0, 1, or 2 leptons (muons or electrons) and two identified \cPqb\ jets. Muons and electrons from \PW\ or \cPZ\ boson decays are identified using the selection criteria defined in~\cite{Sirunyan:2017elk}, including stringent quality and isolation requirements. The lepton \pt requirements are $\pt > 25\,(30)\GeV$ for muons (electrons) in the \onel channel, and $\pt > 20\GeV$ for each lepton in the \twol channel. In the 0- and \onel channels, events with additional isolated muons or electrons are rejected. Minimum \pt requirements of $(60,35)$, $(25,25)$, and $(20,20)\GeV$ are applied on the two jets forming the Higgs boson candidate in the 0-, 1-, and \twol channels, respectively, which are also required to satisfy $\abs{\eta}< 2.5$.

The reconstruction of vector boson decays differs by channel. In the \zerol channel, \ptmiss is interpreted as the \pt of the \cPZ\ boson, while in the \twol channel the \cPZ\ boson is reconstructed directly from lepton pairs requiring an invariant mass in the range 75--105\GeV. In the \onel channel, the \PW\ boson candidate is reconstructed from the \ptvec of the single isolated lepton and \ptvecmiss.

Higgs boson candidates are reconstructed from the pair of jets (``jj'') in the event most likely to originate from \cPqb\ quarks, as determined by a combined secondary vertex algorithm (deepCSV) based on a deep neural network discriminant (DNN)~\cite{Sirunyan:2017ezt}. The deepCSV algorithm provides a continuous discriminator score combining information about tracks displaced from the primary vertex, identified secondary vertices, jet kinematic variables, and information related to the presence of soft leptons in the jet. Of the two jets forming the Higgs boson candidate, the one with a larger deepCSV score is required to satisfy a tight working point with misidentification rate of 0.1\% for light quark and gluon jets, while the jet with a lower deepCSV score (DCSV2) must satisfy a loose working point with a 10\% misidentification rate.

All backgrounds are substantially reduced by requiring large \ptV~\cite{PhysRevLett.100.242001}. The requirements are $\ptmiss>170\GeV$ and $\ptV>150\GeV$ in the 0- and \onel channels, respectively, while two regions are used in the \twol channel: $50 < \ptV \le 150\GeV$ and $\ptV > 150\GeV$. Residual backgrounds from \ttbar and QCD processes are reduced in the \onel channel by rejecting events with more than one extra jet satisfying $\pt > 30\GeV$ and $\abs{\eta}<2.5$, and in the \zerol channel with additional requirements on the angular separation between \ptvecmiss and its nearest jet.

After all event selection criteria are applied, the resolution on the dijet invariant mass \mjj of reconstructed Higgs boson decays is approximately 15\%. The mass resolution is improved by applying a multivariate regression technique similar to that employed in~\cite{Sirunyan:2017elk}, with a DNN trained on \cPqb\ jets from simulated \ttbar events with input variables that include several properties of any secondary vertices in the jet, as well as the energy and composition of the jet. In addition, recovery of final-state radiation is achieved by adding to the \mjj calculation the momenta of jets near to either of the Higgs boson candidate jets. In the \twol channel, with no genuine \ptvecmiss from the hard-scattering process, a kinematic fit of the entire event is performed requiring \pt balance between the dilepton and dijet systems within the experimental uncertainty. All three improvements are validated in data by studying the \ptll/\ptjj distribution in samples of $\cPZ\to\ell\ell$ events containing at least one \cPqb-tagged jet, and by studying the top quark mass distribution in a high-purity sample of \ttbar events. After these improvements, the average resolution on \mjj is in the 10--13\% range, depending on the channel and the \pt of the reconstructed Higgs boson.

For each channel, a signal region enriched in \VH events is selected together with several control regions, each enriched in events from individual background processes. The signal regions are defined as $60<\mjj<160\GeV$ in the \zerol channel, and $90<\mjj<150\GeV$ in the $1$- and \twol channels. The score of a DNN for events in each of these signal regions, which further separates signal from background, is used in the signal extraction fit. The DNNs are trained separately for each channel using simulated samples for signal and all background processes. The set of input variables is chosen by an iterative optimization procedure from a large number of potentially discriminating variables. Among the most discriminating variables for all channels are \mjj, \ptV, DCSV2, the number of additional jets, and the angular separation between the two jets forming the Higgs boson candidate. Events in control regions are used in the fit to normalize the major background processes directly from data. These regions are selected for \ttbar production (TT), and for the production of \PW\ and \cPZ\ bosons in association with either predominantly heavy-flavor (HF) or light-flavor (LF) jets using the criteria described in~\cite{Sirunyan:2017elk}.

The signal strength $\mu$, defined as the measured production cross section times branching fraction divided by the expected SM value, is extracted from a simultaneous binned fit of the signal and control regions. The DNN score is used as the fitted variable in each signal region, while different strategies are used in the control regions. For the TT and LF control regions, only the yields of these processes are considered in the fit. For the HF control region, DCSV2 is used as the fitted variable in the \twol channel, while the score of a dedicated DNN (DNNHF) is used in the 0- and \onel channels. The DNNHF uses the same variables as the signal region DNN, but is trained to individually distinguish the \ttbar, single \cPqt, and \Vjets background processes.

The significance of the observed excess of events in the fit is computed using the profile likelihood asymptotic approximation~\cite{junkcls,Read:2002hq,LHC-HCG-H-no-h,Cowan:2010js}. All results reported here are obtained for a Higgs boson mass of $\mH =125.09\GeV$~\cite{Aad:2015zhl}. For the 2017 data, the observed significance is 3.3$\sigma$ above the background-only hypothesis, while $3.1\sigma$ is expected for the SM Higgs boson. The corresponding measured signal strength is $\mu = 1.08\pm 0.34$, where the uncertainty is a combination of statistical and systematic components. Table~\ref{tab:systematics} lists the major sources of uncertainty and their observed impact on $\mu$ from the fit. Dominant systematic uncertainties arise from the background normalizations, simulated sample size, \cPqb-tagging efficiency and misidentification rates, and \Vjets modeling. All sources of systematic uncertainty are included in the fit as independent nuisance parameters.

\begin{table}[htb]
\centering
\topcaption{Major sources of uncertainty in the measurement of the signal strength $\mu$, and their observed impact ($\Delta\mu$) from a fit to the 2017 data set, are listed. The total uncertainty is separated into four components: statistical (including data yields), experimental, MC sample size, and theory. Detailed decompositions of the statistical, experimental, and theory components are specified. The impact of each uncertainty is evaluated considering only that source. Because of correlations in the combined fit between nuisance parameters in different sources, the sum in quadrature for each source does not in general equal the total uncertainty of each component.}
\begin{scotch}{ l l l }
Uncertainty source & \multicolumn{2}{c}{$\Delta\mu$} \\
\hline
  Statistical                                   & $+0.26$ & $-0.26$ \\
  \qquad Normalization of backgrounds           & $+0.12$ & $-0.12$ \\[\cmsTabSkip]
  Experimental                                  & $+0.16$ & $-0.15$ \\
  \qquad $\cPqb$-tagging efficiency and misid   & $+0.09$ & $-0.08$ \\
  \qquad \Vjets\ modeling                       & $+0.08$ & $-0.07$ \\
  \qquad Jet energy scale and resolution        & $+0.05$ & $-0.05$ \\
  \qquad Lepton identification                  & $+0.02$ & $-0.01$ \\
  \qquad Luminosity                             & $+0.03$ & $-0.03$ \\
  \qquad Other experimental uncertainties       & $+0.06$ & $-0.05$ \\[\cmsTabSkip]
  MC sample size                                & $+0.12$ & $-0.12$ \\[\cmsTabSkip]
  Theory                                        & $+0.11$ & $-0.09$ \\
  \qquad Background modeling                    & $+0.08$ & $-0.08$ \\
  \qquad Signal modeling                        & $+0.07$ & $-0.04$ \\[\cmsTabSkip]
  Total                                         & $+0.35$ & $-0.33$ \\
\end{scotch}
\label{tab:systematics}
\end{table}

The \VZ process with \ZBB, having an identical final state as the \VH process with \HBB, serves to validate the methodology used in the search for the latter process. To extract this diboson signal, the DNNs are trained using the simulated samples for this process as signal. All other processes, including \VH production (at the predicted SM rate), are treated as background. The only modification made to the analysis is the requirement that the signal region is in the interval $[60,160]\GeV$ in \mjj for all channels. The excess of events for the combined \PW\cPZ\ and \cPZ\cPZ\ production processes has an observed significance of $5.2\sigma$ from the background-only hypothesis, where $5.0\sigma$ is expected. The corresponding observed signal strength is $\mu = 1.05\pm 0.22$.

Measurements of the \VH process with \HBB reported above are combined with the results of a similar measurement performed by the CMS Collaboration using data collected at 13\TeV in 2016 corresponding to 35.9\fbinv~\cite{Sirunyan:2017elk}. All systematic uncertainties are assumed to be uncorrelated in this fit, except for theory uncertainties and the dominant uncertainties in the measurement of the jet energy scale, which are assumed to be fully correlated. The Run 2 (2016 and 2017 data sets) combination yields an observed signal significance of $4.4\sigma$, where $4.2\sigma$ is expected, and a signal strength of $\mu=1.06\pm 0.26$.

\begin{figure*}[htb!]
\centering
    \includegraphics[height=0.4\textwidth]{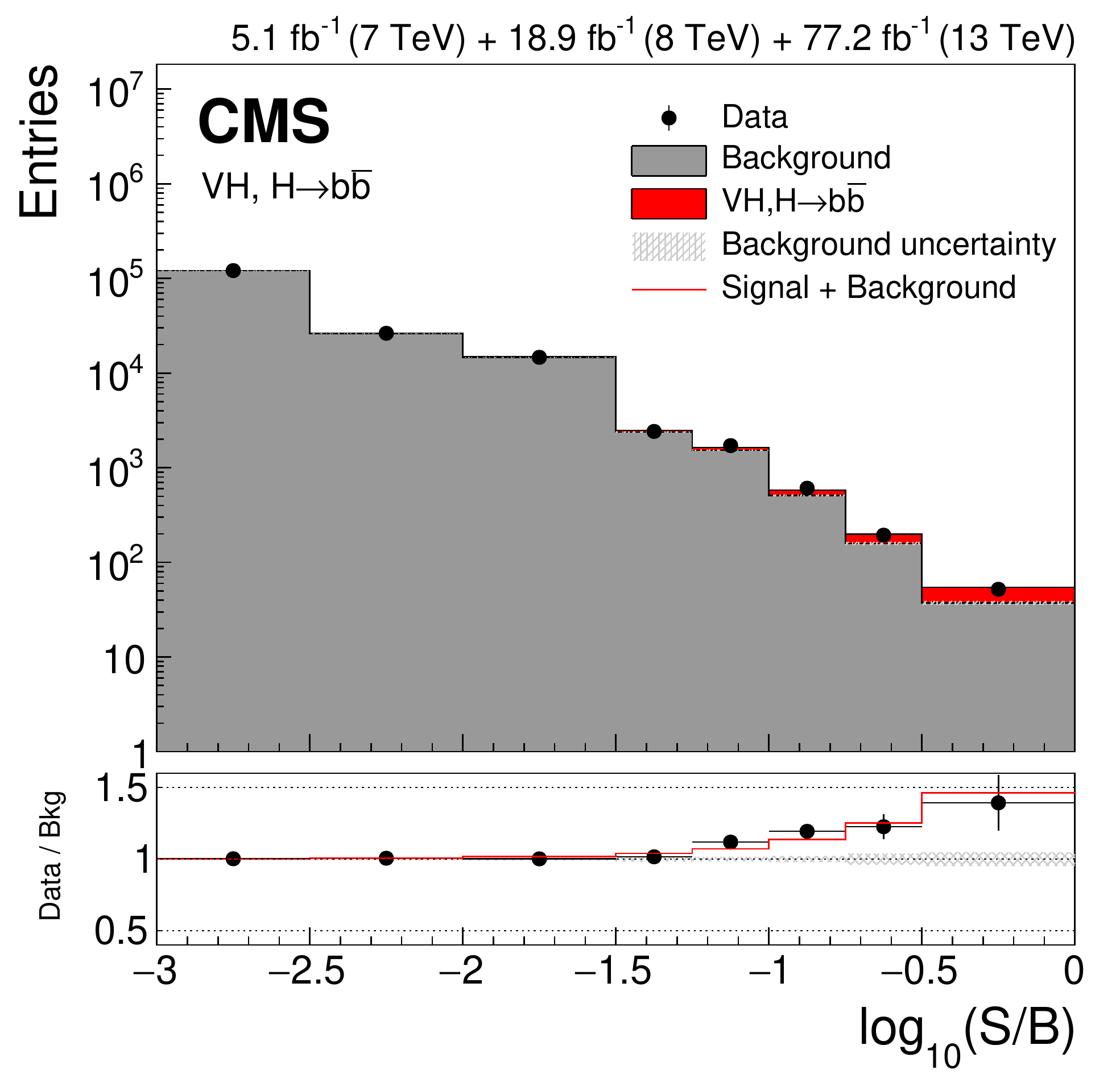}
    \includegraphics[height=0.4\textwidth]{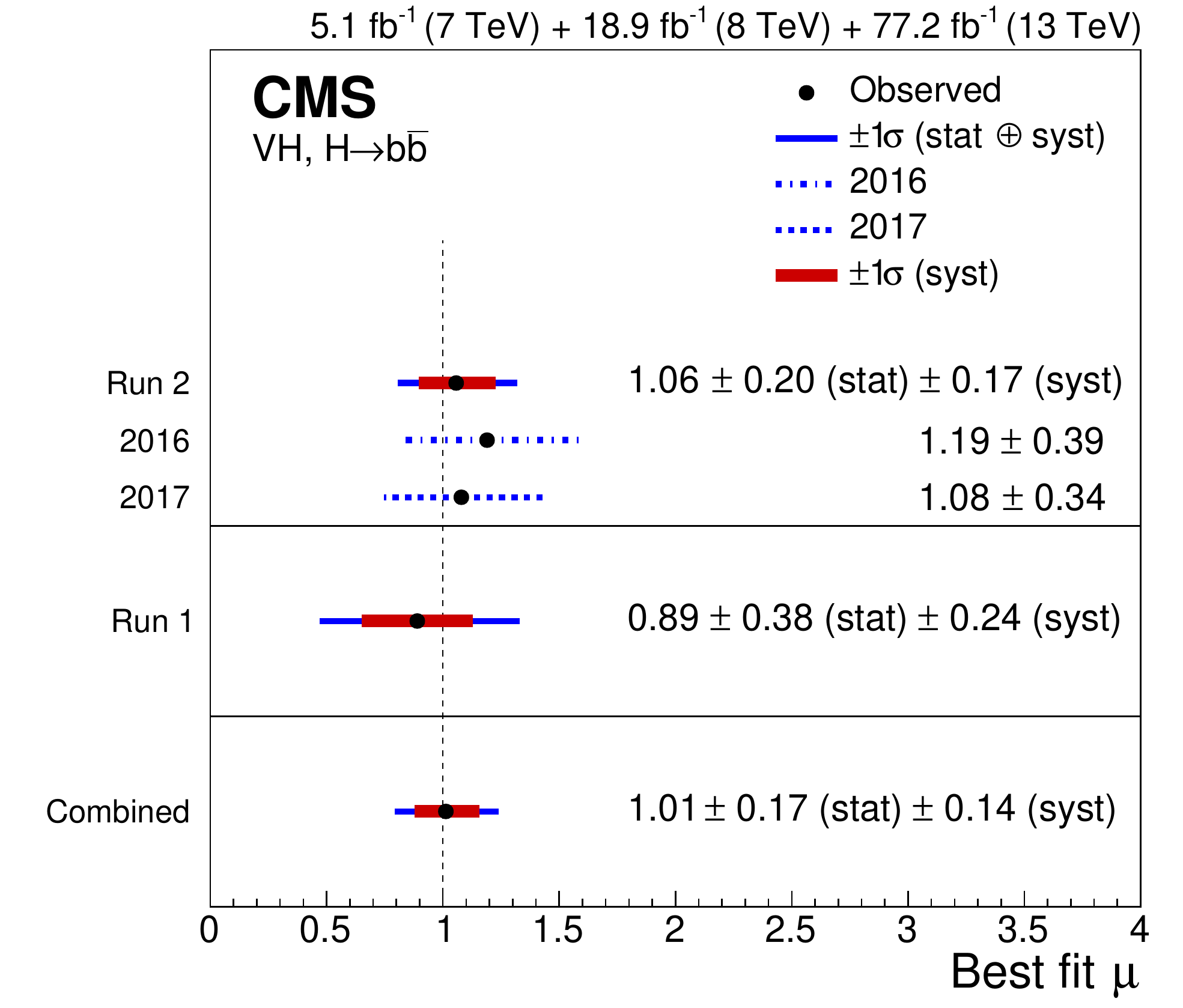}
\caption{Left: distributions of signal, background, and data event yields sorted into bins of similar signal-to-background ratio, as given by the result of the fit to their corresponding multivariate discriminant. All events in the \VH, \HBB signal regions of the combined Run 1 and Run 2 data sets are included. The red histogram indicates the Higgs boson signal contribution, while the grey histogram is the sum of all background yields. The bottom panel shows the ratio of the data to the background, with the total uncertainty in the background yield indicated by the grey hatching. The red line indicates the sum of signal plus background contribution divided by the background yield. Right: best-fit value of the signal strength $\mu$, at $\mH = 125.09\GeV$, for the fit of all \VH, \HBB channels in the Run 1 and Run 2 data sets. Also shown are the individual results of the 2016 and 2017 measurements, the Run 2 combination, and the Run 1 result. Horizontal error bars indicate the $1\sigma$ systematic (red) and $1\sigma$ total (blue) uncertainties, and the vertical dashed line indicates the SM expectation.}
\label{fig:logSB_mjj}
\end{figure*}

The results \VH from Run 2 are combined with the results of a similar CMS analysis of the Run 1 data using $\Pp\Pp$ collisions at $\sqrt{s}=7$ and 8\TeV with data samples corresponding to integrated luminosities of up to 5.1 and 18.9\fbinv, respectively~\cite{Khachatryan:2015bnx,Khachatryan:2016vau}. Systematic uncertainties in this fit are assumed to be uncorrelated for separate collision energies, except for the theory uncertainties. The combination yields an observed signal significance of $4.8\sigma$, where $4.9\sigma$ is expected. The measured signal strength is $\mu = 1.01\pm 0.22\,\left[0.17\stat\pm 0.09\,(\text{exp})\pm 0.06\,(\text{MC})\pm 0.08\thy\right]$, where the decomposition of the total uncertainty into its components is specified in brackets following the definitions in Table~\ref{tab:systematics}. Figure~\ref{fig:logSB_mjj} (left) shows the distribution of events in all channels sorted according to the observed value of $\log_{10}{(S/B)}$ for the combined Run 1 and Run 2 data sets, where signal $S$ and background $B$ yields are determined from the corresponding discriminant score used in each analysis (DNNs for the 2017 data set, boosted decision trees for all other data sets). Figure~\ref{fig:logSB_mjj} (right) summarizes the signal strengths for \VH production, with \HBB, separately for the different data sets and the combination, while Table~\ref{tab:limits_by_mode} summarizes the significances, also including a breakdown of the 2017 results separated by channel.

\begin{table}[htbp]
\centering
\topcaption{Expected and observed significances, in $\sigma$, and observed signal strengths for the \VH production process with \HBB. Results are shown separately for 2017 data, combined Run 2 (2016 and 2017) data, and for the combination of the Run 1 and Run 2 data sets. For the 2017 analysis, results are shown separately for the individual signal strengths for each channel from a combined simultaneous fit to all channels. All results are obtained for $\mH=125.09\GeV$ combining statistical and systematic uncertainties.}
\begin{scotch}{lccc}
                  & \multicolumn{2}{c}{Significance ($\sigma$)}   &                        \\
Data set          & Expected      & Observed           & Signal strength        \\\hline
2017              &               &                    &                        \\
\quad \zerol      & $1.9$         &  $1.3$             & $0.73\pm 0.65$ \\
\quad  \onel      & $1.8$         &  $2.6$             & $1.32\pm 0.55$ \\
\quad  \twol      & $1.9$         &  $1.9$             & $1.05\pm 0.59$ \\
\quad Combined    & $3.1$         &  $3.3$             & $1.08\pm 0.34$ \\[\cmsTabSkip]
 Run 2            & $4.2$         &  $4.4$             & $1.06\pm 0.26$ \\[\cmsTabSkip]
 Run 1 + Run 2    & $4.9$         &  $4.8$             & $1.01\pm 0.22$ \\
\end{scotch}
\label{tab:limits_by_mode}
\end{table}

\begin{figure}[htb!]
\centering
    \includegraphics[width=\cmsFigWidth]{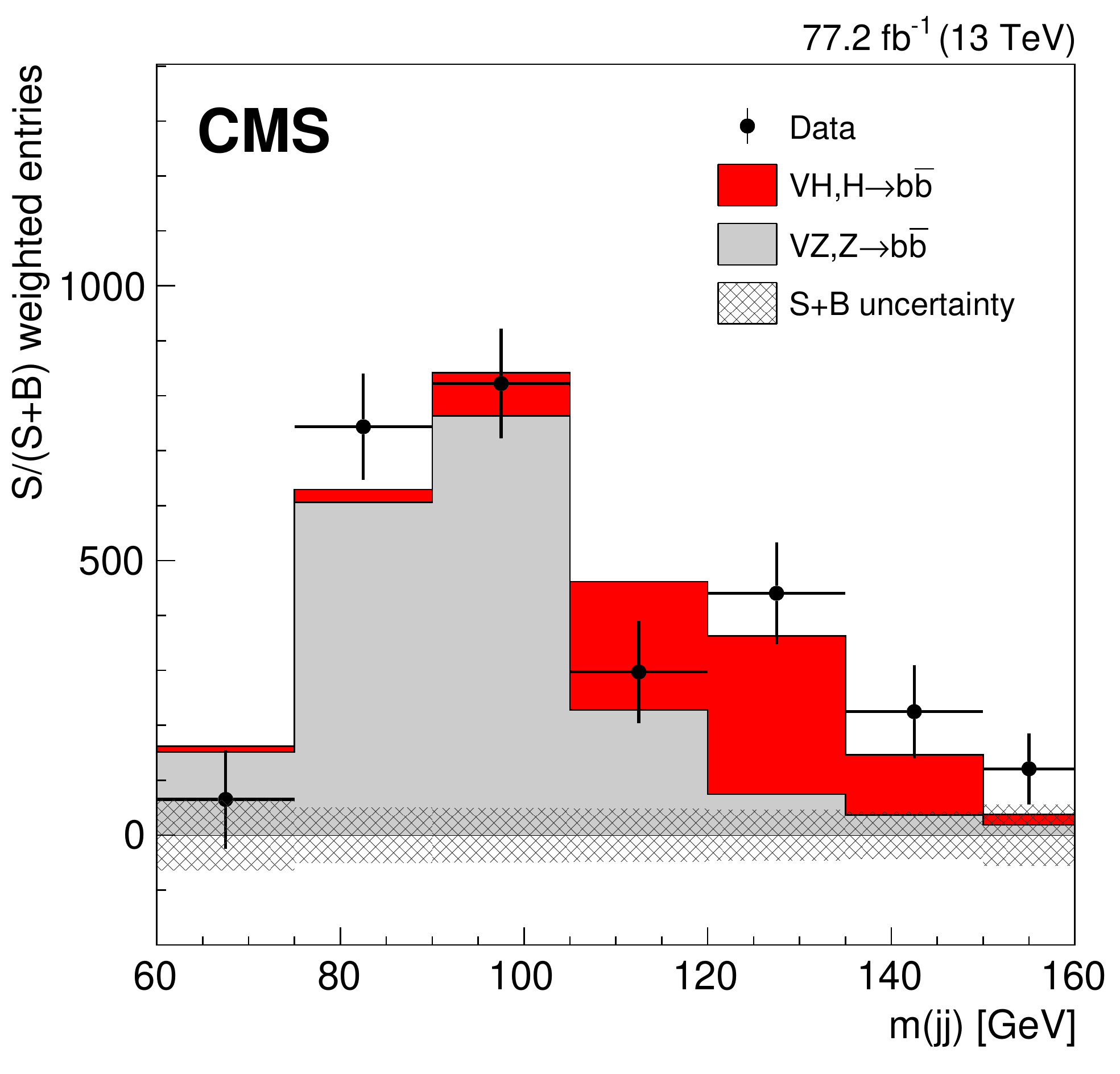}
\caption{Dijet invariant mass distribution for events weighted by $S/(S+B)$ in all channels combined in the 2016 and 2017 data sets. Weights are derived from a fit to the \mjj distribution, as described in the text.  Shown are data (points) and the fitted \VH signal (red) and \VZ background (grey) distributions, with all other fitted background processes subtracted. The error bar for each bin represents the pre-subtraction $1\sigma$ statistical uncertainty on the data, while the grey hatching indicates the $1\sigma$ total uncertainty on the signal and all background components.}
\label{fig:mjj}
\end{figure}

An alternative to fitting the DNN score is to fit the \mjj distribution, which results in less sensitivity but enables a more direct visualization of the Higgs boson signal. As in the \VZ analysis, the signal region is defined to be in the interval $[60,160]\GeV$ in \mjj. This study is performed only with the 2016 and 2017 data sets, in which events are categorized into four bins of increasing signal-to-background ratio according to the score of their corresponding discriminant, obtained with those input variables correlated with \mjj fixed to their mean values. The resulting four \mjj distributions in each data set are fit together with the same distributions used in the control regions, described above, to extract signal and background yields. The fitted \mjj distributions are combined and weighted by $S/(S+B)$, where $S$ and $B$ are computed from the Higgs boson signal yield and the sum of all background yields for each category considering their fitted normalizations,  respectively. The resulting combined \mjj distribution, after background subtraction, is shown in Fig.~\ref{fig:mjj}, where the \VH and \VZ contributions are separately visible.

\begin{figure}[htb!]
\centering
    \includegraphics[width=\cmsFigWidth]{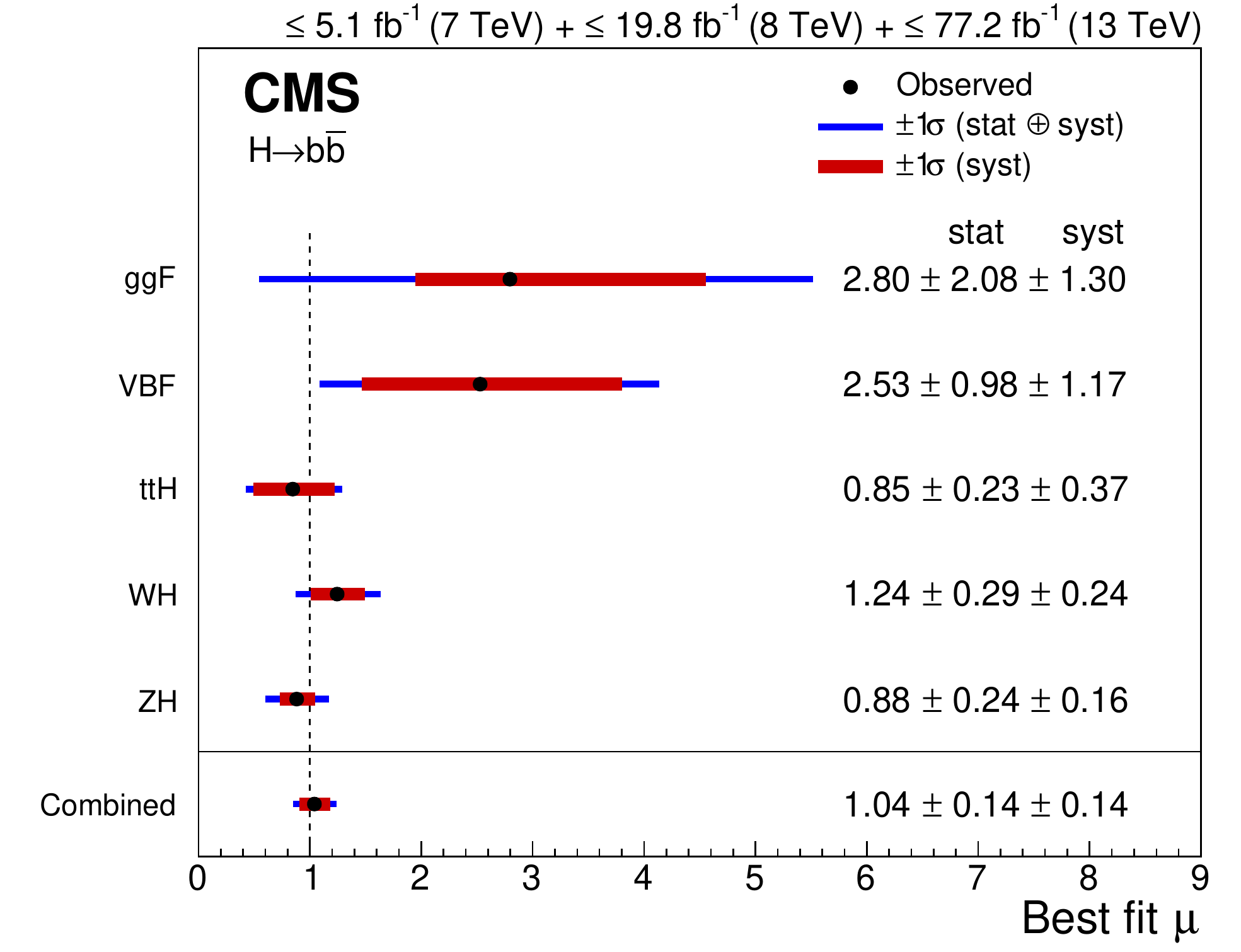}
\caption{Best-fit value of the $\HBB$ signal strength with its $1\sigma$ systematic (red) and total (blue) uncertainties for the five individual production modes considered, as well as the overall combined result. The vertical dashed line indicates the standard model expectation. All results are extracted from a single fit combining all input analyses, with $\mH = 125.09\GeV$.}
    \label{fig:ccplot}
\end{figure}

A combination of CMS measurements of the $\HBB$ decay is performed, including dedicated analyses for the following production processes: \VH (reported above), gluon fusion~\cite{Sirunyan:2017dgc}, vector boson fusion~\cite{Khachatryan:2015bnx}, and associated production with top quarks~\cite{Sirunyan:2018hoz,Chatrchyan:2013yea,Khachatryan:2014qaa}. These analyses use data collected at 7, 8 and 13\TeV, depending on the process. In this fit, most sources of systematic uncertainty are treated as uncorrelated. The dominant jet energy scale uncertainties are treated as correlated between processes at the same collision energy, while the theory uncertainties are correlated between all processes and data sets. The observed (expected) signal significance is 5.6 (5.5)$\sigma$, and the measured signal strength is $\mu=1.04\pm 0.20$. In addition to the overall signal strength for the \HBB decay, the signal strengths for the individual production processes are also determined in this combination, where contributions from a single production process to multiple channels are properly accounted for in the fit. All results are summarized in Fig.~\ref{fig:ccplot}.

In summary, measurement of the standard model Higgs boson decaying to bottom quarks has been presented. A combination of all CMS measurements of the \VH, \HBB process using proton-proton collisions recorded at center of mass energies of 7, 8, and 13\TeV, yields an observed (expected) significance of 4.8 (4.9) standard deviations at $\mH = 125.09\GeV$, and the signal strength is $\mu = 1.01\pm 0.22$.  Combining this result with previous measurements by the CMS Collaboration of the \HBB decay in events where the Higgs boson is produced through gluon fusion, vector boson fusion, or in association with top quarks, the observed\,(expected) significance increases to 5.6\,(5.5) standard deviations and the signal strength is $\mu = 1.04\pm 0.20$. This constitutes the observation of the \HBB decay by the CMS Collaboration.

\begin{acknowledgments}
We congratulate our colleagues in the CERN accelerator departments for the excellent performance of the LHC and thank the technical and administrative staffs at CERN and at other CMS institutes for their contributions to the success of the CMS effort. In addition, we gratefully acknowledge the computing centers and personnel of the Worldwide LHC Computing Grid for delivering so effectively the computing infrastructure essential to our analyses. Finally, we acknowledge the enduring support for the construction and operation of the LHC and the CMS detector provided by the following funding agencies: BMBWF and FWF (Austria); FNRS and FWO (Belgium); CNPq, CAPES, FAPERJ, FAPERGS, and FAPESP (Brazil); MES (Bulgaria); CERN; CAS, MoST, and NSFC (China); COLCIENCIAS (Colombia); MSES and CSF (Croatia); RPF (Cyprus); SENESCYT (Ecuador); MoER, ERC IUT, and ERDF (Estonia); Academy of Finland, MEC, and HIP (Finland); CEA and CNRS/IN2P3 (France); BMBF, DFG, and HGF (Germany); GSRT (Greece); NKFIA (Hungary); DAE and DST (India); IPM (Iran); SFI (Ireland); INFN (Italy); MSIP and NRF (Republic of Korea); MES (Latvia); LAS (Lithuania); MOE and UM (Malaysia); BUAP, CINVESTAV, CONACYT, LNS, SEP, and UASLP-FAI (Mexico); MOS (Montenegro); MBIE (New Zealand); PAEC (Pakistan); MSHE and NSC (Poland); FCT (Portugal); JINR (Dubna); MON, RosAtom, RAS, RFBR, and NRC KI (Russia); MESTD (Serbia); SEIDI, CPAN, PCTI, and FEDER (Spain); MOSTR (Sri Lanka); Swiss Funding Agencies (Switzerland); MST (Taipei); ThEPCenter, IPST, STAR, and NSTDA (Thailand); TUBITAK and TAEK (Turkey); NASU and SFFR (Ukraine); STFC (United Kingdom); DOE and NSF (USA).
\end{acknowledgments}

\bibliography{auto_generated}
\cleardoublepage \appendix\section{The CMS Collaboration \label{app:collab}}\begin{sloppypar}\hyphenpenalty=5000\widowpenalty=500\clubpenalty=5000\vskip\cmsinstskip
\textbf{Yerevan Physics Institute, Yerevan, Armenia}\\*[0pt]
A.M.~Sirunyan, A.~Tumasyan
\vskip\cmsinstskip
\textbf{Institut f\"{u}r Hochenergiephysik, Wien, Austria}\\*[0pt]
W.~Adam, F.~Ambrogi, E.~Asilar, T.~Bergauer, J.~Brandstetter, M.~Dragicevic, J.~Er\"{o}, A.~Escalante~Del~Valle, M.~Flechl, R.~Fr\"{u}hwirth\cmsAuthorMark{1}, V.M.~Ghete, J.~Hrubec, M.~Jeitler\cmsAuthorMark{1}, N.~Krammer, I.~Kr\"{a}tschmer, D.~Liko, T.~Madlener, I.~Mikulec, N.~Rad, H.~Rohringer, J.~Schieck\cmsAuthorMark{1}, R.~Sch\"{o}fbeck, M.~Spanring, D.~Spitzbart, W.~Waltenberger, J.~Wittmann, C.-E.~Wulz\cmsAuthorMark{1}, M.~Zarucki
% ...
\vskip\cmsinstskip
% ...
1:  Also at Vienna University of Technology, Vienna, Austria\\
% ...
\end{sloppypar}
\end{document}